\documentclass{optica-article}

\journal{opticajournal} 

\articletype{Research Article}

\usepackage{graphicx}
\usepackage{dcolumn}
\usepackage{bm}
\usepackage{verbatim}
\usepackage{soul}
\usepackage{braket}
\usepackage{dsfont}
\usepackage{amsmath,subfigure,float}
\usepackage{url}
\usepackage{ctable}
\usepackage{soul}
\usepackage[USenglish]{babel}
\usepackage{titlesec}
\addto{\captionsUSenglish}{}
\usepackage{siunitx}
\DeclareSIUnit\unitpi{\mathrm{\pi}}

\newcommand{\rr}{\mathbf{r}}
\newcommand{\ff}{\mathbf{f}}
\newcommand{\qq}{\mathbf{q}}

\newcommand{\dd}{\textup{d}}

\begin{document}

\title{Synthetic Quantum Holography with Undetected Light}

\author{Sebastian T{\"o}pfer,\authormark{1,$\dagger$,*} Sergio Tovar,\authormark{1,*} Josué R. León Torres,\authormark{2,3,4} Daniel Derr,\authormark{1} Enno Giese,\authormark{1} Jorge Fuenzalida,\authormark{1,5} and Markus Gr{\"a}fe\authormark{1,2,$\ddagger$}}

\address{\authormark{1}Institute for Applied Physics, Technical University of Darmstadt, Schloßgartenstraße 7, 64289 Darmstadt, Germany\\
\authormark{2}Fraunhofer Institute for Applied Optics and Precision Engineering IOF, Albert-Einstein-Str. 7, 07745 Jena, Germany\\
\authormark{3}Abbe Center of Photonics, Friedrich Schiller University Jena, Albert-Einstein-Straße 6, 07745 Jena, Germany\\
\authormark{4}Cluster of Excellence Balance of the Microverse, Friedrich Schiller University Jena, Jena, Germany\\
\authormark{5}Present address: ICFO-Institut  de  Ciencies  Fotoniques,  The  Barcelona  Institute  of Science  and  Technology,  08860  Castelldefels  (Barcelona),  Spain\\
\authormark{*}These authors contributed equally}

\email{\authormark{$\dagger$}sebastian.toepfer@tu-darmstadt.de}
\email{\authormark{$\ddagger$}markus.graefe@tu-darmstadt.de} 

\begin{abstract*}
Utilizing nonlinear interferometers for sensing with undetected light enables new sensing and imaging techniques in spectral ranges that are difficult to detect. To enhance this method for future applications, it is advantageous to extract both amplitude and phase information of an object. This study introduces two approaches for synthetic quantum holography with undetected light, which allows for obtaining an object's amplitude and phase information in a nonlinear interferometer by capturing only a single image. One method is based on quasi-phase-shifting holography using superpixel structures displayed on a spatial light modulator. The other method relies on synthetic off-axis holography implemented through a linear phase gradient on a spatial light modulator. Both approaches are experimentally analyzed for applicability and compared against available multi-acquisition methods.

\end{abstract*}
\section{Introduction}

In the realm of optical imaging, accessing both amplitude and phase information is crucial for a comprehensive understanding of complex phenomena~\cite{park_quantitative_2018}. Holographic imaging is vital in this context, providing detailed optical characterization that is essential for studying materials and biological samples~\cite{gabor1948new}. However, traditional methods are predominantly limited to the visible spectral range, often leaving critical phenomena in challenging spectral domains such as the far infrared largely unexplored due to technological constraints in detection capabilities~\cite{bianconi_recent_2020, li_recent_2023}.  

As recently demonstrated, nonlinear interferometers~\cite{chekhova_nonlinear_2016,hochrainer2022review,fuenzalida2024nonlinear} allow not only for quantum imaging~\cite{lemos_quantum_2014, gilaberte_basset_video-rate_2021,kviatkovsky2020microscopy}, spectroscopy~\cite{kalashnikov2016infrared,lindner2020fourier,tashima2024ultra}, and optical coherence tomography~\cite{valles2018optical,vanselow2020frequency} but also for quantum holography with undetected light (QHUL)~\cite{topfer_quantum_2022,haase2023phase}. These techniques are based on the effect of induced coherence~\cite{zou1991induced}. While this significantly enhanced the capabilities of performing holography in extreme spectral ranges, it still rests upon multiple acquisitions, which are not suitable for capturing dynamic processes that evolve rapidly. To overcome these obstacles, we present an advanced scheme called synthetic QHUL. Our approach allows capturing amplitude and phase information in a single acquisition. This single-shot capability is particularly transformative for observing dynamic processes, allowing for accurate imaging of transient events that traditional multi-acquisition techniques might miss.

Our technique, which ingeniously combines nonlinear interferometers with synthetic holography schemes, has the potential to transform imaging in spectral regions that were traditionally hindered by detector inefficiencies. This simplifies the holographic imaging process and enhances the practicality of quantum imaging applications in real-world scenarios, benefiting from partial robustness against environmental noise~\cite{fuenzalida_experimental_2023}.
The manuscript provides an experimental analysis of two synthetic QHUL techniques that allow  phase and transmission acquisition. By building on the innovative trajectory of prior quantum imaging research, our work represents a significant leap forward in using nonlinear interferometers to overcome the challenges of traditional optical imaging, thereby opening up new and groundbreaking applications across scientific and technological fields.

\begin{figure}[bp]
    \centering
    \includegraphics[width=1\textwidth]{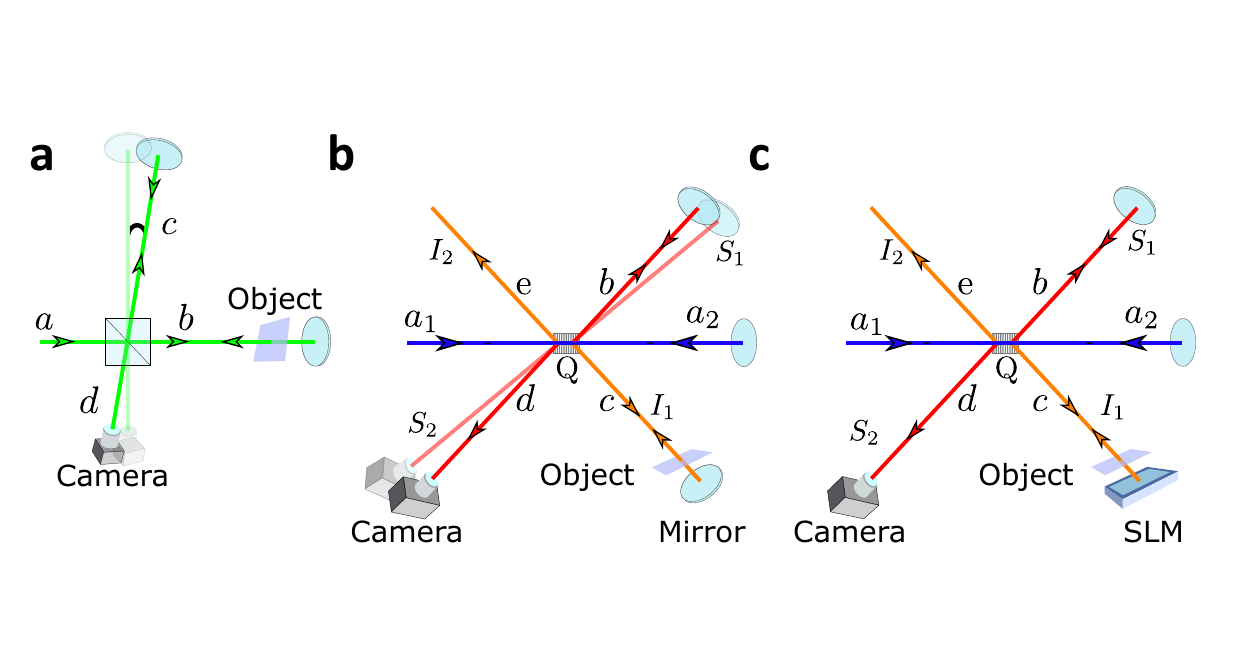}
    \caption{\textbf{Single-acquisition holography schemes.} \textbf{a} Classical off-axis holography scheme based on a Michelson interferometer. Due to the tilt of one of the beam paths, the camera will record an interference pattern that allows the extraction of amplitude and phase information of the object. \textbf{b} Off-axis QHUL implemented in an SU(1,1) interferometer. Photons with different wavelengths traverse different paths in the interferometer, and the object illumination spectrally differs from the detection on the camera. The tilting mirror is placed in one of the end mirrors. The photon that illuminates the object remains undetected. The information of the object is transferred to its partner photon as coherence. \textbf{c} Synthetic quantum holography scheme. In contrast to \textbf{a} and \textbf{b}, no active tilting of any end mirror is necessary. Due to the nonlinear interferometry and the spatial light modulator (SLM), an interference pattern is recorded, allowing a single acquisition to extract amplitude and phase in an undetected light scheme without loss of indistinguishability.
    }
    \label{fig:Setup}
\end{figure}
\section{Principle}

A classical technique for single acquisitions is off-axis holography, which is schematically depicted in Fig.~\ref{fig:Setup}\textbf{a}. The end mirror is tilted in one of the Michelson interferometer arms, resulting in a spatially dependent phase difference between the two arms. By Fourier-analysis-dependent post-processing, the amplitude and phase information can be extracted. 

Fig.~\ref{fig:Setup}\textbf{b} shows the implementation of the same technique for a nonlinear interferometer, called off-axis QHUL. In this arrangement, photons with different wavelengths traverse three interferometric arms. The tilting mirror is placed in one of the end mirrors, producing the desired interference patterns. This also introduces an unwanted loss of indistinguishability and therefore visibility. The object's transmission and phase information can be read out from the light beam that does not interact with it. Fig.~\ref{fig:Setup}\textbf{c} sketches the implementation of synthetic QHUL. An SLM is placed in one of the arms to apply the desired synthetic holographic technique. It displays a linear phase gradient for synthetic off-axis QHUL or super-pixel structures for quasi-phase-shifting QHUL. Both synthetic techniques benefit from intrinsic robustness and reduction of experimental complexity since no moving or tilting components are necessary. Both methods and their implementation are discussed in sections~\ref{Sec:off-axis} and~\ref{Sec:superpixel}, respectively.  
\section{Experimental Setup}
\label{chap:theory}

\begin{figure}[bp]
    \centering
    \includegraphics[width=1\textwidth]{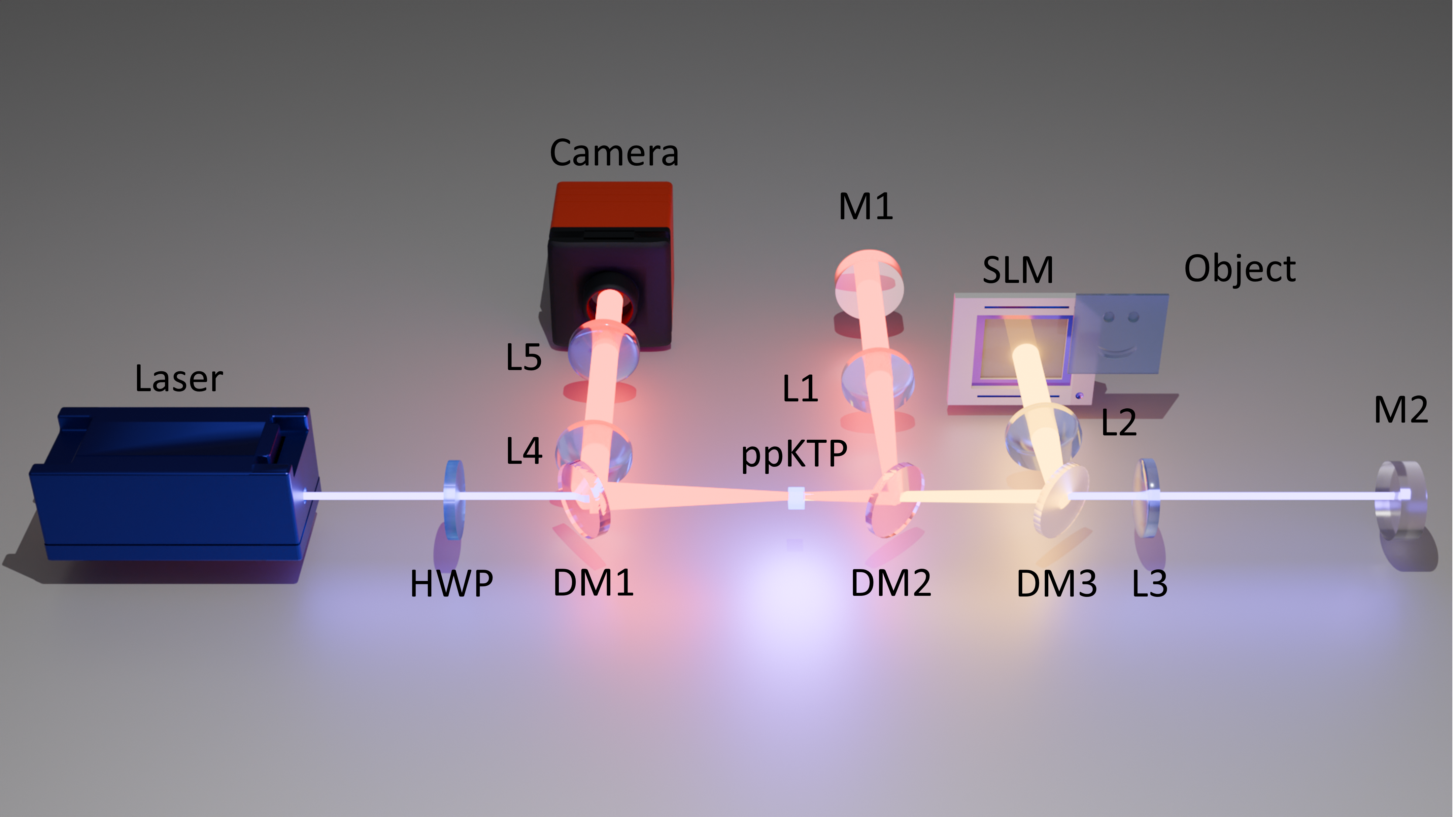}
    \caption{\textbf{Experimental setup.} The experimental setup uses a \SI{405}{\nm} pump laser to generate SPDC photons with a periodically poled potassium titanyl phosphate (ppKTP) crystal. The dichroic mirrors DM1 to DM3 split the different wavelengths into the interferometer arms. The lenses L1 to L3, altogether with the end mirrors M1, M2, and the SLM, form 4-f optical systems to create beam overlap for the forward and backward traveling SPDC beams. This ensures path indistinguishability. The object is placed in the idler arm in front of the SLM. According to the synthetic holography scheme, the SLM provides a position-dependent phase shift of the impinging beam. The lenses L4 and L5 image the object plane into the camera, resulting in an interference image with the object information being recorded.}
    \label{fig: Setup}
\end{figure}

The experiment was carried out using an SU(1,1) nonlinear interferometer
~\cite{yurke_su2_1986} shown in Fig.~\ref{fig: Setup}. Photon pairs are generated in a quantum superpositions of two propagation directions: forward and backward. 
For this purpose, a pump laser (Toptica top-mode with a wavelength of $\lambda_p=\SI{405}{\nm}$  and a pump power of \SI{90}{\mW} impinges a periodically poled potassium titanyl phosphate (ppKTP) crystal for a spontaneous parametric down-conversion (SPDC) generation. The down-converted photons are usually called signal (s) and idler (i). The half-wave plate matches the polarization of the pump photons for the SPDC process. Considering a type-0 phase-matching SPDC process, the biphoton quantum state in the transverse momentum representation is given by~\cite{walborn_spatial_2010}
\begin{align}
    \ket{\psi}_{\mathrm{SPDC}}= \int \dd \qq_\mathrm{i} \, \dd \qq_\mathrm{s} \,  C(\qq_\mathrm{i},\qq_\mathrm{s}) \, \ket{\qq_\mathrm{i}} \ket{\qq_\mathrm{s}}  ,
\label{Eq:bi-photon-SPDC}
\end{align}
where $C$ is the joint amplitude function of signal and idler modes, and $\ket{\qq_\mathrm{i}}$ ($\ket{\qq_\mathrm{s}}$) represents a single-photon state of the idler (signal) photon occupying the transverse wave vector $\qq_\mathrm{i}$ ($\qq_\mathrm{s}$). In this work, the wavelengths of the signal and idler photons are $\lambda_s=$\SI{910}{\nm} and $\lambda_i=$\SI{730}{\nm}, respectively. We also use the naming convention established in the quantum imaging with undetected light community, i.e., we consider the idler as the undetected photon.

    Inside the SU(1,1) interferometer, pump, signal, and idler beams are spatially separated with two dichroic mirrors (DM). While DM1 and DM2 reflect the signal photon, DM3 reflects the idler photon. The lenses L1, L2, and L3 with the same focal length of $f_i=f_s=f_p=$ \SI{100}{\mm} perform a Fourier transform of the near-field plane (crystal plane) on the signal, idler, and pump arms, respectively.
    In the resulting far-field plane in the idler arm, an object with a complex transmission coefficient $|T(\rr_\mathrm{O})| \exp \{\mathrm{i}\phi_\mathrm{T}(\rr_\mathrm{O})\}$ is placed followed by a SLM (Holoeye Pluto NIR\;118 with \SI{8}{\micro \meter} pixel size). 
    Thus, idler light with a transverse momentum $\hbar \qq_\mathrm{i}$ illuminates a point $\rr_\mathrm{O}$ on the object, where $\rr_\mathrm{O}\approx f_i \lambda_i \qq_i/(2\pi)$~\cite{fuenzalida_resolution_2022}. The SLM reflects the idler beam back to the crystal.
    The SLM also plays a fundamental role in our experiments since it allows the realization of different operations on the idler beam transverse phase.
    As shown in~\cite{herzog_frustrated_1994} the three wavelengths of the idler, signal, and pump photons are related, and it would also be possible to place the SLM in the signal or pump end-mirrors to apply the phase changes, making this implementation flexible. In general, the SLM can introduce a spatially dependent phase, $\exp \{\mathrm{i}\phi_{\mathrm{SLM}}(\rr_\mathrm{O})\}$, that will be used to introduce synthetic holographic techniques. The SLM allows us to deploy and compare different holographic techniques without the need for mechanically moving elements or changing the optical path of the back reflected idler beam, resulting in higher stability and preservation of the amplitude's interference~\cite{zou1991induced,hochrainer2017interference}. In addition, we employ a spatially independent phase ($\phi_{\mathrm{SLM}}(\rr_\mathrm{O})\rightarrow\phi_{\mathrm{SLM}}$) to perform digital phase-shifting (DPS) QHUL~\cite{topfer_quantum_2022}. Because this work will compare different QHUL techniques, we will further on refer to this specific as DPS QHUL. This technique will serve as a reference for the synthetic QHUL techniques presented in this work. In the other arms of the SU(1,1) interferometer, the signal and pump beams are reflected back to the crystal by mirrors M1 and M2, respectively. On their way back, the three beams traverse the respective lenses again, producing 4-f systems. This is essential to produce indistinguishability between forward and backward SPDC processes. Forward and backward-generated biphotons are aligned, and DM1 reflects the signal light toward the camera
    The lens L4 with a focal length of 100 mm performs a Fourier transform of the crystal plane at its back focal plane, and the lens L5 with also a focal length of 100 mm images this plane into the camera's plane.
    On the contrary, the idler photon remains undetected. Due to the imaging system and the momentum correlations between signal and idler photons, the information on one point of the object $\rr_\mathrm{O}$ is transferred to one point on the camera $\rr_\mathrm{C}$, with $\rr_\mathrm{C}\approx f_C \lambda_s \qq_s/(2\pi)$, $f_C$ is the focal length that performs the Fourier transform multiple by any additional magnification of the optical system from the crystal to the camera. The signal photon intensity at one point on the camera is 
\begin{align}
    I(\rr_\mathrm{C}) = \int \dd \qq_\mathrm{i} P(\qq_\mathrm{i}, \qq_\mathrm{s}) \{1+\,|T(\rr_\mathrm{O})| \, \cos[
    \phi_\mathrm{T}(\rr_\mathrm{O})
    + \phi_{\mathrm{SLM}}(\rr_\mathrm{O})]\}.
    \label{Eq:signal.with.obj.}
\end{align}
where $P$ is the joint probability, $P(\qq_\mathrm{i},\qq_\mathrm{s})=|C(\qq_\mathrm{i},\qq_\mathrm{s})|^2$, which determines the degradation of the images spatial resolution, due to imperfect momentum correlations. \cite{fuenzalida_resolution_2022}, c.f. \cite{viswanathan_resolution_2021, vega_fundamental_2022, gilaberte_basset_experimental_2023}.

\section{Synthetic off-axis quantum holography with undetected light}\label{Sec:off-axis}
\subsection{Method}
\begin{figure}[bp]
	\centering
\includegraphics[width=\textwidth]{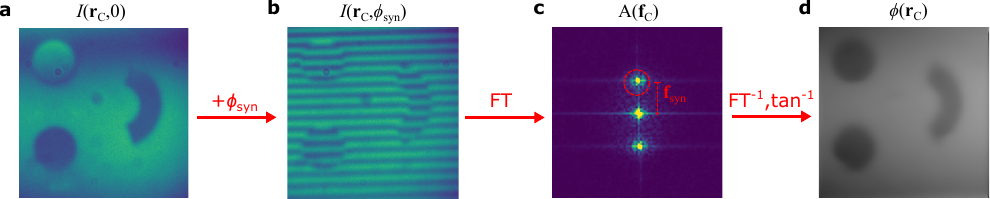}
	\caption{\textbf{Synthetic off-axis QHUL post-processing.} The image shows the post-processing steps for synthetic off-axis QHUL. \textbf{a} Recorded intensity image without applied phase $\phi_\mathrm{SLM}$. \textbf{b} Recorded intensity image after applying a linear phase of shape $\phi_\mathrm{SLM} \propto y_\mathrm{O}/\lambda_\mathrm{syn}$. \textbf{c} Fourier transformation of \textbf{b} with the auto-correlation term in the center and first-order cross-correlation terms above and below separated due to the applied phase. \textbf{d} Calculated phase after applying an inverse Fourier transformation on only the upper cross-correlation term that is marked by a red dashed circle in \textbf{c}.}
	\label{Fig:soff_post_processing}
\end{figure}

The idea of synthetic off-axis QHUL combines nonlinear interferometer with synthetic optical holography based on synthetic reference waves as presented in~\cite{schnell_synthetic_2014}. Fig.~\ref{Fig:soff_post_processing} shows the experimental steps and the post-processing to combine these techniques. Fig.~\ref{Fig:soff_post_processing}\textbf{a} displays an interference image taken by the camera of the nonlinear interferometer. To apply synthetic optical holography to simulate off-axis holography, the SLM displays a linear phase of the shape 
\begin{align}
    \phi_\mathrm{SLM}(y_\mathrm{O}) &= 2\pi \frac{y_\mathrm{O}}{\lambda_\mathrm{syn}}
\end{align}
with $y_\mathrm{O}$ as the vertical component of $\rr_\mathrm{O}$ and $\lambda_\mathrm{syn}$ as the wavelength of the synthetic reference wave. As a consequence, the recorded intensity images will show a sine pattern typical for off-axis holography (Fig.~\ref{Fig:soff_post_processing}\textbf{b}), creating a synthetic image-plane hologram. This will lead to a shift of the cross-correlation terms in the Fourier domain shown in Fig.~\ref{Fig:soff_post_processing}\textbf{c} (see Supplemental Material section~1 for detailed information). It is important that the wavelength $\lambda_\mathrm{syn}$ leads to a clear separation of the cross-correlation term from the auto-correlation term in the center, but still within the Field-of-View. Using a Gaussian window function and moving one of the cross-correlation terms to the center as indicated in Fig.~\ref{Fig:soff_post_processing}\textbf{c} will have the inverse Fourier transformation to contain the separated object information. In the experiments, the center points for the Fourier image $A(\ff_\mathrm{C})$ were removed completely before moving the cross-correlation term because the Gaussian window function was not lowering values of this area sufficiently. The object information will be retrieved as complex amplitude; therefore, the visibility of the interference (and, with this, the object transmission) can be obtained by applying the absolute on the re-transformed images and the phase information by using the arc tangent of the imaginary part divided by the real part as shown in Fig.~\ref{Fig:soff_post_processing}\textbf{d}. The obtained phase images are subtracted by a reference phase afterwards. The reference is a phase image recorded without sample. This allows to obtain only phase change introduced by the sample. Similarly, the transmission images are divided by a reference transmission image to only obtain the transmission of the sample itself.
\subsection{Results and Discussion}

\begin{figure}[bp]
    \centering
    \includegraphics[width=\textwidth]{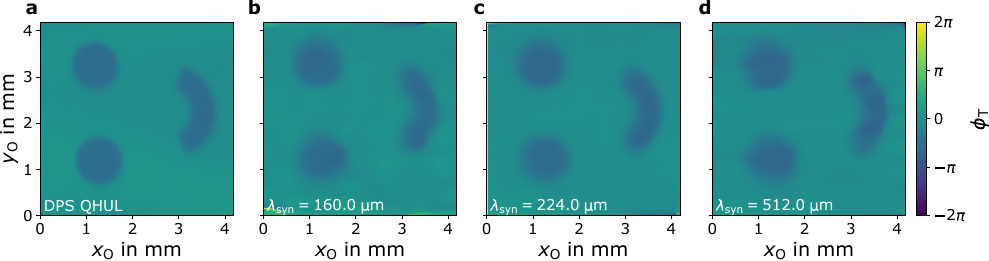}
    \caption{\textbf{Phase images}. \textbf{a} Phase image created using digital phase shifting (DPS) QHUL with four phase steps. \textbf{b} - \textbf{d} Phase images created using synthetic off-axis QHUL with phase gradients to generate vertical sine pattern with \SI{160.0}{\micro m}, \SI{224.0}{\micro m} and \SI{512.0}{\micro m} wavelengths, respectively. }
    \label{fig:phases_compilation}
\end{figure}

\subsubsection{Phase imaging}
\begin{figure}[htbp]
    \centering
    \includegraphics[width=0.9\textwidth]{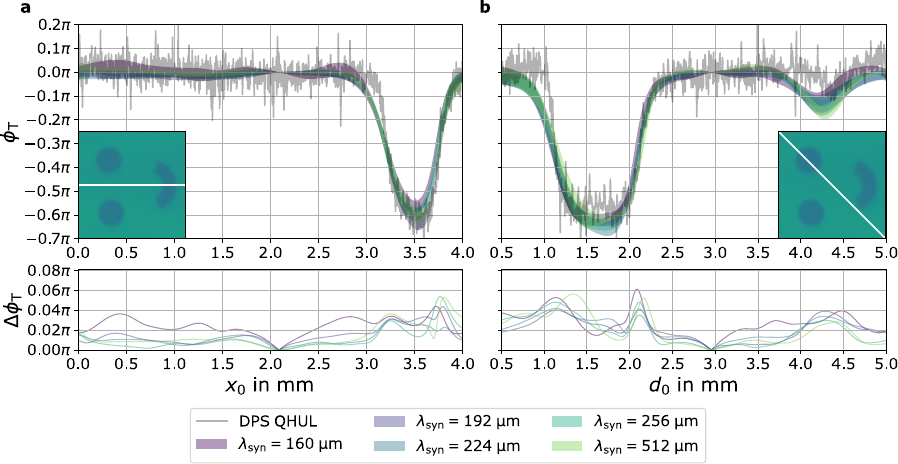}
    \caption{\textbf{Cuts through phase images comparing phase values obtained with DPS QHUL and synthetic off-axis QHUL}.
    The top plots show the phase values of DPS QHUL (gray trace) and synthetic off-axis QHUL for different synthetic wavelengths $\lambda_\text{syn}$ (colored shading) along \textbf{a} an horizontal and \textbf{b} an diagonal cut parameterized by the coordinate $d_\mathrm{O} = \sqrt{x_\mathrm{O}^2 + y_\mathrm{O}^2}$ (see insets for reference). The DPS QHUL values show the mean of the retrieved phases featuring high-spatial-frequency modulations, while the synthetic off-axis QHUL values are displayed by the area of plus and minus one standard deviation around their mean.
    The bottom panels display the standard deviations of synthetic off-axis QHUL and show their position dependence.}
    \label{fig:cut_phases}
\end{figure}

Fig.~\ref{fig:phases_compilation} shows a selection of the measured object phases for the DPS QHUL and synthetic off-axis QHUL.
The happy face object was inscribed into a glass plate via grey-scale lithography.
The phases were calculated as described above and recorded with an exposure time of \SI{120}{\ms} and a visibility of \num{0.33 \pm 0.01}. While the object's shape is correctly retrieved, the spatial resolution is reduced,
as expected from a Gaussian window function in Fourier space.
For a more quantitative analysis, we average over 10 consecutive phase images. 
Two different cuts through these phase images, one along the horizontal and one along the diagonal direction, are shown in Fig.~\ref{fig:cut_phases}.
The top part compares the phase retrieved for different synthetic wavelengths $\lambda_\mathrm{syn}$, where the shaded area marks the standard deviation obtained from the averaging process.
We see that they all show a similar behavior and capture the phase steps of the object.

From the object geometry, we expect a phase difference of \SI{0.62}{\unitpi}, which is within the error margin of our recordings. We measured a phase difference of \SI{0.58 \pm 0.05}{\unitpi} and \SI{0.61 \pm 0.03}{\unitpi} with DPS QHUL and synthetic off-axis QHUL respectively.

Next, we compare the extracted phases to ones obtained from DPS QHUL along the same cuts (without showing its standard deviation).
While the shape of the object and the height of the phase steps generally agree, we observe three major differences:

(i) The mean value obtained from DPS QHUL fluctuates between different positions.
We attribute this behavior to two effects, namely, that for DPS QHUL different recordings enter the algorithm so that the technique is prone to pump laser fluctuations and there is no spatial Gaussian filter applied, which would in principle suppress high spatial-frequency fluctuations, but also deteriorate the detection of edges.

(ii) The edges recorded by DPS QHUL are sharper than for synthetic off-axis QHUL, while to lowest order the slope detected with synthetic off-axis QHUL does not depend on $\lambda_\text{sync}$.
This behavior is a consequence of applying a Gaussian window function of width $\sigma = \SI{6.0}{\per\mm}$ in Fourier space.
This width was chosen independently of $\lambda_\text{sync}$ and in such a way that the side peaks arising from the pixelation of the SLM are always suppressed.
In fact, the processed signal is the convolution of a Gaussian in position space with the interference pattern, which itself includes the point spread function of the imaging system and deviations from perfect anti-correlations of signal and idler photons induced by a finite angular profile of the pump.
While the edge detection of DPS QHUL is only prone to the latter effects, the width of the synthetic off-axis QHUL is additionally broadened by the Gaussian filter.
Similarly, the cut along the diagonal (shown in Fig.~\ref{fig:cut_phases}\textbf{b}) features a dip of \SI{-0.1}{\unitpi} around \SI{3}{\mm}.
While the object in fact has no phase step at this point, the cut is almost tangent to a phase feature of the object. 
Hence, due to the limited spatial resolution of synthetic off-axis QHUL, it is also observed in our phase recordings.
As expected, this feature is suppressed for DPS QHUL.

(iii) Only one side of the detected phase features shows a deterioration in the edge gradient.
This behavior is an artefact originating from the limited accuracy in shifting the cross-correlation term in the Fourier space during the post-processing for synthetic off-axis QHUL (see Supplemental Material section 2), given by the finite pixel size in Fourier space limited by the dimensions of the recorded images.

We furthermore compare the phase sensitivity observed for detection with different $\lambda_\mathrm{syn}$ in the bottom of Fig.~\ref{fig:cut_phases} by plotting the respective standard deviation obtained from averaging 10 phase images.
As a first observation, we note that the standard deviation depends on the pixel position on the camera and is in general in a similar level, even though the shortest wavelength $\lambda_\mathrm{syn}= 160$\,\textmu m is considerably larger.
As a second feature, we observe that the standard deviation increases at the edges of the phase object, which is not surprising since the phase uncertainty depends on the phase itself, as outlined in the Supplemental Material.
And finally, we observe that around the center of the Fourier transform, i.\,e., the center of the image, all variances decrease linearly independent of the wavelength.
This effect can be derived explicitly from the expressions for the phase uncertainty in the Supplemental Material and is a result of the pixelation of Fourier space and the associated imperfect shifts.

\subsubsection{Transmission images}

\begin{figure}[bp]
    \centering
    \includegraphics[width=\textwidth]{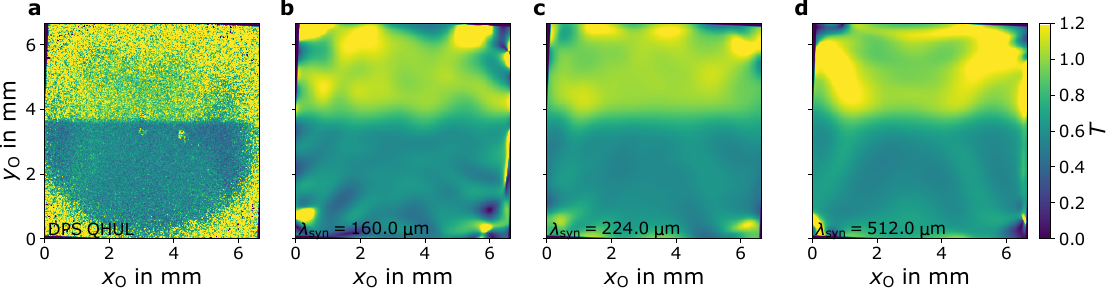}
    \caption{\textbf{Transmission images}. \textbf{a} Transmission image created using digital phase shifting (DPS) QHUL with four phase steps. \textbf{b} - \textbf{d} Transmission images created using synthetic off-axis QHUL with phase gradients to generate vertical sine pattern with \SI{160.0}{\micro m}, \SI{224.0}{\micro m} and \SI{512.0}{\micro m} wavelengths, respectively.}
    \label{fig:transmission_compilation}
\end{figure}
\begin{figure}[htbp]
    \centering
    \includegraphics[width=0.9\textwidth]{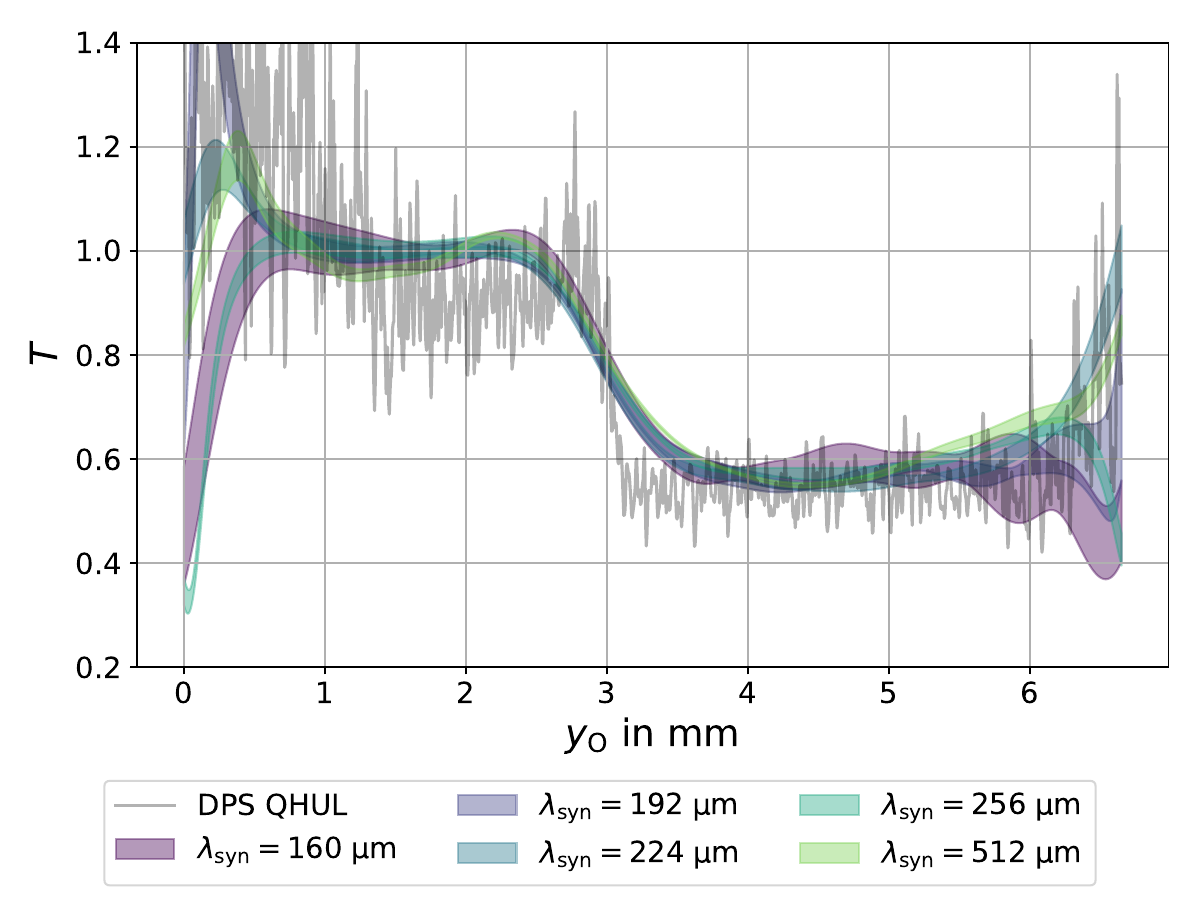}
    \caption{\textbf{Comparison of transmission values obtained with DPS QHUL and synthetic off-axis QHUL}. The plot shows the transmission values of DPS QHUL (gray trace) and synthetic off-axis QHUL for different synthetic wavelengths $\lambda_\text{syn}$ (colored shading) along a vertical cut through the image center. The DPS QHUL values show the mean of the retrieved transmission featuring high-spatial-frequency modulations, while the synthetic off-axis QHUL values are displayed by the area of plus and minus one standard deviation around their mean.}
    \label{fig:cut_transmission}
\end{figure}

Fig.~\ref{fig:transmission_compilation} compares the obtained transmission using the edge between two areas of a variable neutral density filter as a sample (NDL-25S-2 from Thorlabs, Inc.). The transmission step was taken between regions of optical density \num{0.3} and \num{0.6}. Therefore, a transmission step of \num{0.5} is expected. We measured a transmission step of \num{0.52 \pm 0.16} and \num{0.59 \pm 0.12} with DPS QHUL and synthetic off-axis QHUL, respectively. Therefore, the correct value was retrieved within the uncertainty.
Compared to Fig.~\ref{fig:phases_compilation} the image of the DPS QHUL show more noise, due to a loss in dynamic range of the camera. 
The compensate the effect to certain degrees, the acquisition time was increased to \SI{300}{\ms}, giving a visibility of \num{0.30 \pm 0.08}. Nonetheless the maximum Intensity was lower with $\num{340\pm10}$ compared to $\num{650\pm7}$ arb. camera units, resulting into higher relative fluctuations.

Fig.~\ref{fig:cut_transmission} shows the obtained values. along a vertical cut through the image center. Four major effects can be seen:

(i) Similar to the phase values in Fig.~\ref{fig:cut_phases}, the mean values for the transmission obtained by DPS QHUL show visible fluctuations. With the same reasoning we see the contributions of the intensity fluctuations of the pump beam and the absence of a gaussian filter in the fourier space as cause.   

(ii) On the outsides of the shown region, the mean values as well as the standard deviations for both methods increase significantly. This regions are at the border or outside the illuminated area on the sample. As a consequence the obtained reference values, used for division are approaching zero causing this effect, due to the mathematical operation.

(iii) Just as seen for the phase values in Fig.~\ref{fig:cut_phases}. The spatial structure of the edge between the different transmission levels is broadened compared to the values obtained with synthetic off-axis QHUL, due to the use of a Gaussian window function. The spatial width $\sigma$ of the window function was \SI{4.5}{\per\mm}.

(iv) Another interesting effect can be seen in the spatial structure of the transmission values acquired with $\lambda_\mathrm{syn} = \SI{512}{\micro \meter}$ in Fig,~\ref{fig:transmission_compilation}. Overtone frequencies are visible, leading to a falsification of the mean values obtained. Among the compared values for $\lambda_\mathrm{syn}$, this one has the lowest distance of the cross-correlation term from the central auto-correlation term in the Fourier space. Due the reduced dynamic range of the camera, noise in the frequency sidebands of the auto-correlation term is more pronounced. It was not possible to completely remove this in the post-processing, because to the width of the Gaussian window function. 

\subsection{Discussion}

The results show that the synthetic off-axis QHUL can measure phase values and transmission values with comparable accuracy as DPS QHUL. Because of a window function used in the image post-processing process, it will induce a loss in spatial resolution. But because only one frame is needed, it was possible to achieve a speed increase of four times. Further improvements can be reached by optimizing the window function used in synthetic off-axis QHUL to  suppress frequency side-bands more efficiently, while lowering the width of the filter to reduce overtones in the spatial structure obtained and increase the spatial resolution.
We measured a time increase in the post-processing of approximately factor four compared to DPS QHUL (see Supplemental Material section 3). As the post-processing can be done separately from the acquisition, the presented method allows for the recording of dynamic imaging scenes, as a consequence of the low total acquisition time needed. It must be mentioned that the DPS QHUL can also be applied with three interference frames at the cost of reduced accuracy \cite{topfer_quantum_2022}, which still suggests a speed increase of three times.


\section{Superpixel quantum holography with undetected light}\label{Sec:superpixel}
\subsection{Method}

Another alternative approach to DPS holography is 
called parallel quasi-phase shifting digital holography and was introduced in~\cite{awatsuji_parallel_2004}. The idea is to use a structured light beam to induce the phase shifts in parallel. It consists of dividing the object beam into pixels. A group of four pixels with the phases $0$, $\pi/2$, $\pi$, and $3\pi/2$ is called a superpixel. The superpixel is then repeated to tessellate the entire object beam.

\begin{figure}[htbp]
\centering
\includegraphics[scale = 0.6]{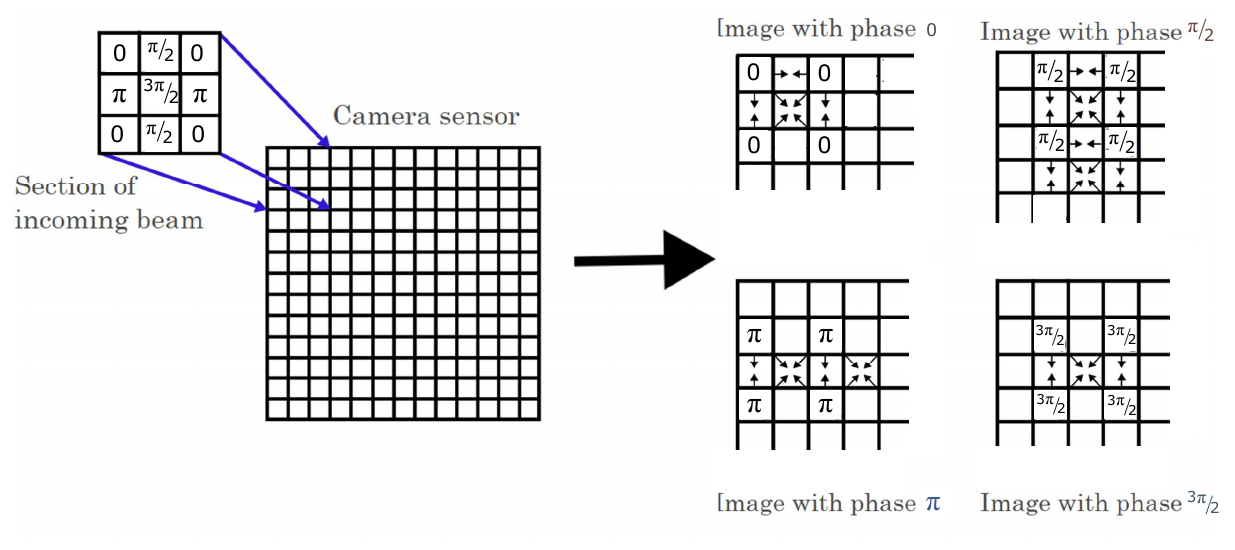}
\caption{\textbf{Working principle of parallel quasi-phase-shifting.} The phase-shifted beam arrives at the camera sensor. Only a section of it is depicted. The whole image detected at the sensor is split into four separate sets. Each set contains an intensity image that has been phase-shifted under a given phase step. Each of the four images is completed by interpolating the holes using the neighboring available values. The phase is calculated according to DPS QHUL.}
\label{fig:Recon}
\end{figure}

Transferring this approach into the quantum holography domain, we introduce a superpixel structure on the SLM. Each superpixel is formed by 2$\times$2 subpixels where the subpixels have the relative phases to each other of $0$, $\pi/2$, $\pi$, and $3\pi/2$. Due to the induced coherence effect in the nonlinear interferometer, the pixelization of the beam is present in the signal beam impinging on the camera.
As depicted in Fig.~\ref{fig:Recon}, with one camera frame, four sub-images with the corresponding relative phases between them can be obtained and used for DPS QHUL. As before, the total resolution of the imaging system is reduced in exchange for a fourfold reduction in acquisition time.

\subsection{Results and discussion}
\begin{figure}[htbp]
	\centering
    \includegraphics[width=0.8\linewidth]{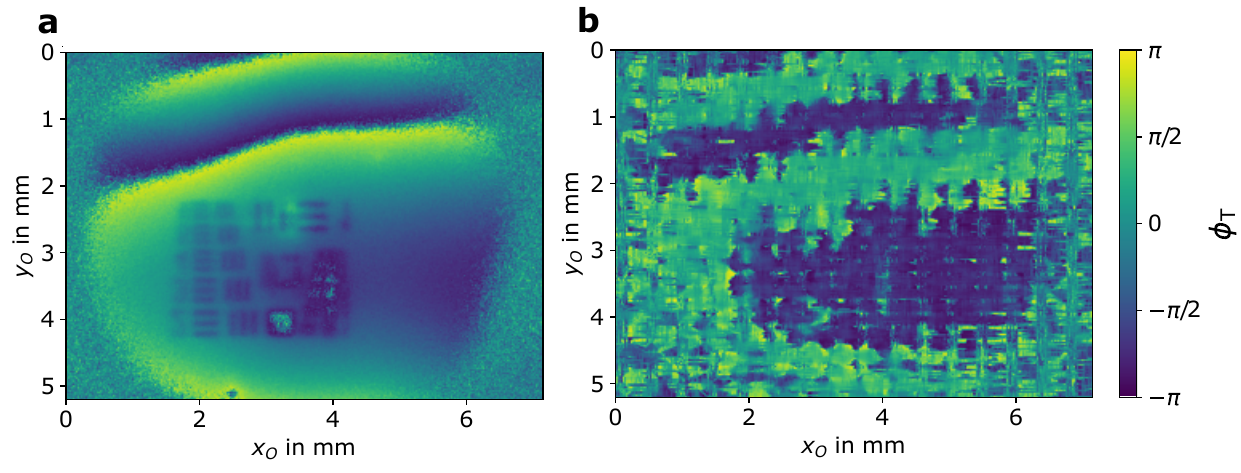}
	\caption{\textbf{Comparison between DPS QHUL and synthetic superpixel QHUL.} The image shows the phase measurements of a miniaturized USAF target fabricated by grey-scale lithography in a glass plate. \textbf{a} Using DPS QHUL. \textbf{b} Using the synthetic superpixel QHUL.}
	\label{Fig:soff_post_processing1}
\end{figure}

Two phase measurements are shown in Fig.~\ref{Fig:soff_post_processing1}. A miniaturized USAF target engraved in a glass plate was used as the object for both phase measurements. The exposure time was \SI{120}{\ms}. According to Fig.~\ref{Fig:soff_post_processing1}\textbf{a} in the DPS QHUL method, the interference rings and the general morphology of the USAF target can be resolved. Although numbers cannot be resolved, general structures can still be distinguished. However, for measurement with the synthetic superpixel QHUL method, all information about the object has been lost, as seen in Fig.~\ref{Fig:soff_post_processing1}\textbf{b}. Only information about the largest phase features of the measurement can be resolved.

In the case of the synthetic superpixel QHUL approach, the resolution of the setup is not only limited by the optics, as in the case of the DPS technique, but also by the SLM pixel size and the superpixel itself. This size is at minimum four pixels to display all phases required for the DPS technique. Assuming a deviation from a one to one mapping of the SLM pixels onto the camera pixels, padding might additionally be needed to prevent cross-talk. In our case these conditions led to a size of the superpixels, which was 320\;$\mu$m. This, in turn, resulted in the loss of features smaller than this superpixel size.

Furthermore, some artifacts arise from this technique. Such artifacts can be seen in Fig.~\ref{Fig:soff_post_processing1}\textbf{b}, the measurement is pixelated, revealing not only the size of the superpixel but also bright horizontal lines that correspond to the unsuccessful selection and interpolation of the individual phase images. Hence, this method is highly dependent on both the execution of the technique and the minimal size of the superpixel.

Further study to improve the separation of the phases would proof crucial to increase the accuracy of the retrieved phase (and transmission) values. Additionally to ensuring a one-to-one mapping of the SLM pixel with the camera pixels, research if a weighing by overlap of the single pixels contribution into the interpolated phase images might also lead to better results.

The post-processing of this technique took around four times longer compared to the DPS QHUL (see supplement material section 3). The post-processing can be done after the acquisition of all the images, therefore it does not inhibit the possibility to measure dynamic scenes after improving the image resolution significantly.

\section{Conclusion}
In this work, we have introduced a pioneering technique of synthetic quantum holography with undetected light, representing a significant advancement in quantum imaging. This novel method combines the principles of nonlinear interferometers and quantum imaging with undetected light together with synthetic holography to capture both amplitude and phase information of an object in a single shot, thereby addressing the limitations posed by traditional multi-exposure methods. While writing our manuscript, we found two publications addressing single-shot QHUL~\cite{leon-torres_off-axis_2024, pearce_single-frame_2024}. Our work fundamentally differs from those by exploiting \emph{synthetic} QHUL, benefiting from single acquisitions and a highly robust setup without moving parts.
Our experimental results confirm the practicality of synthetic off-axis QHUL, suggesting its ability to accurately capture dynamic processes and transient phenomena, which would otherwise be elusive with slower, multi-shot techniques. Despite a decrease in spatial resolution compared to traditional QHUL methods, the gain in imaging speed—up to four times faster is a substantial improvement, particularly useful in real-time and \textit{in vivo} applications where low acquisition time is crucial. Exploiting super-pixel structures as an alternative method features poor resolution for practical purposes at the moment.
Overall, the integration of synthetic holography into the broader spectrum of quantum imaging techniques not only enhances our understanding of complex (quantum) optical phenomena but also broadens the scope of their practical applications, pushing the boundaries of what is visually and scientifically achievable.

\section{Acknowledgement}
Support is acknowledged from the German Federal Ministry of Education and Research (BMBF) within the funding program “quantum technologies – from basic research to market” with contract number 13N16496 (QUANCER). We also acknowledge financial support from Sony Semiconductor Solutions Europe and thank Markus Kamm and Dr. Alexander Gatto for fruitful discussions.

\section{Disclosure}
The authors declare no conflicts of interest.

\section{Data availability}
Data underlying the results presented in this paper are not publicly available at this time but may be obtained from the authors upon reasonable request.

\bibliography{main}

\end{document}